# On four-dimensional Einsteinian gravity, quasitopological gravity, cosmology and black holes

Adolfo Cisterna[a], Nicolás Grandi[b], Julio Oliva[c]

[a]Universidad Central de Chile, Vicerrectoría académica,
*Toesca 1783 Santiago, Chile*

[b]Instituto de Física de La Plata - CONICET & Departamento de Física - UNLP,
*C.C. 67, 1900 La Plata, Argentina.*

[c]Departamento de Física, Universidad de Concepción,
*Casilla 160-C, Concepción, Chile.*

**Abstract**

We show that the combination of cubic invariants defining five-dimensional quasitopological gravity, when written in four dimensions, reduce to the version of four-dimensional Einsteinian gravity recently proposed by Arciniega, Edelstein & Jaime, that produces second order equations of motion in a FLRW ansatz, with a purely geometrical inflationary period. We introduce a quartic version of the four-dimensional Einsteinian theory with similar properties, and study its consequences. In particular we found that there exists a region on the space of parameters which allows for thermodynamically stable black holes, as well as a well-defined cosmology with geometrically driven inflation. We briefly discuss the cosmological inhomogeneities in this setup. We also provide a combination of quintic invariants with those properties.

## I. INTRODUCTION

Any ultra-violet completion of General Relativity will generically induce higher curvature modifications to the space-time dynamics. From a bottom-up point of view, a gravity theory containing higher order terms in the curvature must be restricted by some basic health conditions. Examples of such are the non-existence/decoupling of ghost modes around maximally symmetric vacua, suitable phenomenological restrictions on the scalar mode of the graviton, positivity of the effective Newton's constant, as well as existence of black holes with reasonable thermodynamics. This has fuelled a lot of activity in the area in the last few years (see e.g. [1]-[11], and references therein).

Recently, the authors of [12] proposed a cubic action in which the scalar and massive graviton acquire an infinite mass, and the relative coefficients of the cubic curvature terms are fixed in a dimension-independent manner. This theory also leads to simple spherically symmetric black holes, characterized by a single metric function which satisfies a third-order equation that admits a first integral [13, 14]. The couplings can be chosen so that arbitrarily small black holes are thermodynamically stable, in the local sense [13]. The effects on lensing and black hole shadows have been explored in [15] and [16]. As recently discovered in [17], improving the theory with a second, ghost-free, cubic combination in four dimensions, which vanishes on the static black hole ansatz [19], leads to interesting cosmological scenarios. Remarkably, it gives second order differential equations in the Friedmann-Lemaître-Robertson-Walker (FLRW) ansatz, with a purely geometrical inflationary period, smoothly matching a matter dominated era followed by late time acceleration [17]. A point to be stressed is that the sign of the cubic coupling that gives rise to black holes, is the opposite to that originating healthy cosmologies [18].

In a completely different context, a combination of cubic invariants in five dimensions was originally introduced in [20] in order to produce a second order constraint from the trace of the field equations. Furthermore on a spherically symmetric ansatz, the theory has first order equations, leading to a Birkhoff's theorem[1]. In this paper, we show that the quasitopological combination of cubic invariants, when considered in four dimensions, exactly matches the four-dimensional Einsteinian theory of [17].

---

[1] The same combination was dubbed Quasitopological gravity in [21] where some holographic properties of the theory were also studied.

We construct the general quartic combination that has the same properties than it's cubic Einsteinian counterpart, *i.e.* stability around the maximally symmetric background, spherically symmetric black holes defined by a single metric function that satisfies a third order equation with a first integral, and second order equations of motion on the FLRW ansatz. As shown below, the inclusion of the quartic term allows for a region of parameters in which a healthy cosmology exists as well as thermodynamically stable black holes. We also discuss briefly the cosmological scalar perturbations, showing that they have second order equations of motion. Furthermore, we propose a quintic combination with the same properties.

## II. FOUR-DIMENSIONAL CUBIC EINSTEINIAN GRAVITY AND FIVE-DIMENSIONAL QUASITOPOLOGICAL GRAVITY

For arbitrary dimensions, there are eight independent algebraic cubic invariants constructed out from the Riemann tensor [22]. The basis can be chosen as

$$\begin{aligned}
\mathcal{L}_1 &= R_a{}^c{}_b{}^d R_c{}^e{}_d{}^f R_e{}^a{}_f{}^b \,, & \mathcal{L}_2 &= R_{ab}{}^{cd} R_{cd}{}^{ef} R_{ef}{}^{ab} \,, \\
\mathcal{L}_3 &= R_{abcd} R^{abc}{}_e R^{de} \,, & \mathcal{L}_4 &= R_{abcd} R^{abcd} R \,, \\
\mathcal{L}_5 &= R_{abcd} R^{ac} R^{bd} \,, & \mathcal{L}_6 &= R_a{}^b R_b{}^c R_c{}^a \,, \\
\mathcal{L}_7 &= R_{ab} R^{ab} R \,, & \mathcal{L}_8 &= R^3 \,.
\end{aligned} \quad (1)$$

In terms of these, the unique quasitopological theory in five dimensions, that leads to second order trace of the field equations, has the following form

$$\mathcal{L}_{QTG} = \frac{1}{5}\left(48 b_8 - \frac{11}{3} b_5\right) \mathcal{L}_1 + \frac{1}{4}(b_5 - 8 b_8) \mathcal{L}_2 + \frac{1}{5}(24 b_8 - 6 b_5) \mathcal{L}_3 + \frac{1}{10}(6 b_8 + b_5) \mathcal{L}_4 \\
+ b_5 \mathcal{L}_5 + \frac{2}{5}(16 b_8 + b_5) \mathcal{L}_6 - \frac{1}{5}(36 b_8 + b_5) \mathcal{L}_7 + b_8 \mathcal{L}_8 \,, \quad (2)$$

where the degeneracy in $b_5$ and $b_8$ comes from the vanishing in five dimensions of the six dimensional Euler density

$$\mathcal{E}_6 = -8\mathcal{L}_1 + 4\mathcal{L}_2 - 24\mathcal{L}_3 + 3\mathcal{L}_4 + 24\mathcal{L}_5 + 16\mathcal{L}_6 - 12\mathcal{L}_7 + \mathcal{L}_8 \,. \quad (3)$$

When $\mathcal{L}_{QTG}$ is considered as a four-dimensional theory, a new degeneracy appears due to the vanishing of the combination

$$R^{\mu\nu} \mathcal{G}^{(2)}_{\mu\nu} = \mathcal{L}_3 - \frac{1}{4}\mathcal{L}_4 - 2\mathcal{L}_5 - 2\mathcal{L}_6 + 2\mathcal{L}_7 - \frac{1}{4}\mathcal{L}_8 \,, \quad (4)$$

where $\mathcal{G}_{\mu\nu}^{(2)}$ is the Euler-Lagrange derivative of the Gauss-Bonnet Lagrangian $\sqrt{-g}(R^2 - 4R_{ab}R^{ab} + R_{abcd}R^{abcd})$. With this, it is straightforward to show that the four dimensional Einsteinian theory of reference [17]

$$\mathcal{P} - 8\mathcal{C} = 12\mathcal{L}_1 + \mathcal{L}_2 - 8\mathcal{L}_3 + 2\mathcal{L}_4 + 4\mathcal{L}_5 + 8\mathcal{L}_6 - 4\mathcal{L}_7 \, , \tag{5}$$

identically coincides with the quasitopological combination

$$\mathcal{P} - 8\mathcal{C} = -\left(\frac{60}{b_5 - 24b_8}\right)\mathcal{L}_{QTG} + 16\, R^{\mu\nu}\mathcal{G}_{\mu\nu}^{(2)} + 4\left(\frac{b_5 - 9b_8}{b_5 - 24b_8}\right)\mathcal{E}_6 \, , \tag{6}$$

since, as mentioned above, in four dimensions $\mathcal{E}_6 \equiv 0$ and $\mathcal{G}_{\mu\nu}^{(2)} \equiv 0$.

### III. FOUR-DIMENSIONAL QUARTIC EINSTEINIAN GRAVITY WITH HEALTHY COSMOLOGIES AND BLACK HOLES

In arbitrary dimensions, there are 26 quartic, algebraically independent contractions of the Riemann tensor [22]. The following terms can be used as a basis

$$\begin{aligned}
& c_1\, R^{pqrs}R_{p\ r}^{\ t\ u}R_{t\ q}^{\ v\ w}R_{uvsw} + c_2\, R^{pqrs}R_{p\ r}^{\ t\ u}R_{t\ u}^{\ v\ w}R_{qvsw} + c_3\, R^{pqrs}R_{pq}^{\ \ tu}R_{r\ t}^{\ v\ w}R_{svuw} \\
& + c_4\, R^{pqrs}R_{pq}^{\ \ tu}R_{rt}^{\ \ vw}R_{suvw} + c_5\, R^{pqrs}R_{pq}^{\ \ tu}R_{tu}^{\ \ vw}R_{rsvw} + c_6\, R^{pqrs}R_{pqr}^{\ \ \ t}R^{uvw}_{\ \ \ \ s}R_{uvwt} \\
& + c_7\, R^{pqrs}R_{pqrs}R^{abcd}R_{abcd} + c_8\, R^{pq}R^{rstu}R_r^{\ vtp}R_{svuq} + c_9\, R^{pq}R^{rstu}R_{rs}^{\ \ vp}R_{tuvq} \\
& + c_{10}\, R^{pq}R_{p\ q}^{\ r\ s}R^{tuv}_{\ \ \ r}R^{tuvs} + c_{11}\, RR^{pqrs}R_{p\ r}^{\ t\ u}R_{qtsu} + c_{12}\, RR^{pqrs}R_{pq}^{\ \ tu}R_{rstu} \\
& + c_{13}\, R^{pq}R^{rs}R_{p\ \ r}^{\ t\ u}R_{tqus} + c_{14}\, R^{pq}R^{rs}R_{p\ \ q}^{\ t\ u}R_{trus} + c_{15}\, R^{pq}R^{rs}R^{tu}_{\ \ pr}R_{tuqs} \\
& + c_{16}\, R^{pq}R_p^{\ r}R^{stu}_{\ \ \ q}R_{stur} + c_{17}\, R^{pq}R_{pq}R^{rstu}R_{rstu} + c_{18}\, RR^{pq}R^{rst}_{\ \ \ p}R_{rstq} \\
& + c_{19}\, R^2 R^{pqrs}R_{pqrs} + c_{20}\, R^{pq}R^{rs}R_r^{\ t}R_{psqt} + c_{21}\, RR^{pq}R^{rs}R_{prqs} + c_{22}\, R^{pq}R_p^{\ r}R_q^{\ s}R_{rs} \\
& + c_{23}\, (R^{pq}R_{pq})^2 + c_{24}\, RR^{pq}R_p^{\ r}R_{qr} + c_{25}\, R^2 R_{pq}R^{pq} + c_{26}\, R^4 \, .
\end{aligned} \tag{7}$$

In four dimensions, the list of algebraically independent quartic invariants reduces to 13 terms. We constrain the relative coefficients $c_i$, in order to have similar properties than those of the four dimensional Einsteinian cubic gravity of [17], namely

- The theory must be ghost-free around the maximally symmetric background.

- A FLRW ansatz must lead to second order equations of motion.

- A static, spherically symmetric ansatz, fulfilling $-g_{tt}g_{rr} = 1$, must lead to third order equations of motion with a trivial first integral.

These restrictions lead to constraints on the coefficients (see Appendix A), that defines our four-dimensional theory. The resulting theory reads

$$I[g] = \int \sqrt{-g} d^4 x \left( \frac{R - 2\Lambda_0}{2\kappa} + \kappa \beta \left( \mathcal{P} - 8\mathcal{C} \right) + \kappa^2 \mathcal{Q} \right), \tag{8}$$

where $\mathcal{Q}$ is the quartic combination of the terms (7) with the coefficients $c_i$ restricted as in Appendix A. Here $\kappa = 8\pi G$ and $\beta$ is a dimensionless couplings.

In black holes, in cosmologies of the FRLW-type, as well as in the effective Newton's constant, only the following combination of quartic coefficients appears

$$\gamma = \frac{1}{15} \left( 56 c_1 + 140 c_2 + 112 c_4 + 224 (c_5 + c_6) + 896 c_7 \right), \tag{9}$$

where $\gamma$ is a dimensionless coupling.

Applying the linearization procedure of [4], around a maximally symmetric background of curvature $\Lambda/3$, one obtains

$$\frac{1}{4\kappa} \left( 1 + \frac{16}{3} \beta \kappa \Lambda^2 - \frac{8}{63} \gamma \kappa^2 \Lambda^3 \right) \Box G^L_{\mu\nu} = \frac{1}{4} T^L_{\mu\nu}, \tag{10}$$

where

$$G^L_{\mu\nu} = \left( R^L_{\mu\nu} - \frac{1}{2} g_{\mu\nu} R^L - \Lambda h_{\mu\nu} \right), \tag{11}$$

$$R^L_{\mu\nu} = \nabla_{(\mu|} \nabla_\sigma h^\sigma_{|\nu)} - \frac{1}{2} \Box h_{\mu\nu} - \frac{1}{2} \nabla_\mu \nabla_\nu h + \frac{4\Lambda}{3} h_{\mu\nu} - \frac{\Lambda}{3} h g_{\mu\nu}, \tag{12}$$

$$R^L = \nabla^\mu \nabla^\nu h_{\mu\nu} - \Box h - \Lambda h. \tag{13}$$

Implying that the effective Newton's constant $\kappa_{eff}$ reads

$$\frac{1}{4\kappa_{eff}} = \frac{1}{4\kappa} \left( 1 + \frac{16}{3} \beta \kappa \Lambda^2 - \frac{8}{63} \gamma \kappa^2 \Lambda^3 \right). \tag{14}$$

In Appendix B, we also provide a quintic combination with the aforementioned properties.

### A. Black holes

On a static spherically symmetric ansatz

$$ds^2 = -N^2(r) f(r) dt^2 + \frac{dr^2}{f(r)} + r^2 \left( d\theta^2 + \sin^2 \theta d\phi^2 \right), \tag{15}$$

the equations of motion of the above defined quartic theory (8) imply $N(r) = 1$, reducing to a single equation for $f(r)$, which reads

$$\frac{1}{3}\Lambda_0 r^6 + mr^3\kappa + r^4(f-1) = 4r\kappa^2\beta \left(f'\left(3f' - 3rff'' + rf'^2\right)r - 6f(1-f)(f''r - f')\right)$$
$$+ \frac{3\kappa^3\gamma}{56}f'\left(\left(3rf'^2 - 12rff'' + 4ff' + 8f'\right)f'r - 24f(1-f)(f''r - f')\right), \quad (16)$$

where we have already performed the trivial first integral, introducing the integration constant $m$. In general, this equation has four branches of solutions. There is an Einstein branch, *i.e.* a solution that matches the Schwarzschild-(A)dS metric for small $\beta$ and $\gamma$ [13]. In it, when the bare cosmological constant $\Lambda_0$ vanishes, the metric must be asymptotically flat and the mass of the spacetime is given by the term with fall-off $m/r$ in the metric function $f(r)$. This is only achieved for $\beta < 0$ [13]. For this regard the sign of the quartic coupling is irrelevant, and arbitrarily small black holes always exist.

The sign of the specific heat of the small black holes is dominated by the highest possible curvature coupling, in our case by $\gamma$. Matching with the sign of the healthy small black holes of reference [13] requires $\gamma < 0$.

Remarkably, as we will see below, the restrictions $\beta < 0, \gamma < 0$ coming from the existence and stability of the black holes, render a healthy cosmology.

### B. Cosmology

#### 1. Cosmological evolution

On a Friedmann-Lemaître-Robertson-Walker ansatz

$$ds^2 = -dt^2 + a^2(t)\left(\frac{dr^2}{1 - kr^2} + r^2 d\Omega_2^2\right), \quad (17)$$

our quartic theory (8) gives rise to a Friedmann equation with the form

$$P\left[H^2 + \frac{k}{a^2}\right] = \frac{\kappa}{3}\rho + \frac{\Lambda_0}{3}, \quad (18)$$

where $P[x]$ is a quartic polynomial on $x$

$$P[x] = x + 16\kappa^2\beta x^3 - \frac{6}{7}\kappa^3\gamma x^4, \quad (19)$$

and the energy density $\rho$ contains the relativistic ($\sim a^{-4}$) and non-relativistic ($\sim a^{-3}$) components. By using continuity equation, Friedmann equation can be linked to its second order form

$$P'\left[H^2 + \frac{k}{a^2}\right]\frac{\ddot{a}}{a} = -\frac{\kappa}{2}(p+\rho) + \left(H^2 + \frac{k}{a^2}\right)P'\left[H^2 + \frac{k}{a^2}\right]. \tag{20}$$

In order for the cosmological solution to exist, the polynomial $P[x]$ must intersect the line $(\kappa\rho + \Lambda_0)/3$. It vanishes at $x = 0$ with $P'[0] = 1$. Using Descartes' sign rule to count the possible numbers of roots of $P'[x]$, and the behavior of $P[x]$ and $P'[x]$ at large $\pm x$, we find that for negative $\beta$ and $\gamma$, there is a region of parameters for which the polynomial has a single negative minimum at negative $x$. Since $\kappa\rho + \Lambda_0$ is strictly positive for positive $\Lambda_0$, the minimum is never reached by the cosmological evolution. These signs of the couplings coincide with those ensuring the existence and thermal stability of small black holes as analyzed in the previous section. In what follows, we concentrate in such region of parameters.

As in the cubic case of [17], the resulting cosmological solutions present an inflationary era in the past, of a purely geometric origin, smoothly connected to a matter dominated epoch, followed by a late time acceleration. Some typical profiles of the scale factor are depicted in Figure 1.

In order to match with cosmological observables in a model independent manner, we start by simply expanding the scale factor in powers of the cosmological time as

$$a(t) = a_0\left(1 + H_0 t - \frac{H_0^2 q_0}{2}t^2 + \frac{H_0^3 j_0}{6}t^3 + \frac{H_0^4 s_0}{24}t^4 + \cdots\right), \tag{21}$$

in terms of the kinematic parameters known as decelleration $q_0$, cosmological jerk $j_0$ and snap $s_0$. Together with the spatial curvature, these are purely geometric quantities that can in principle be measured in a model-independent way. Now, we plug this expansion in the Friedmann equation, rewritten in the form

$$P\left[H_0^2\left(1 - \frac{\Omega_k^0}{a^2}\right)\right] = H_0^2\left(\frac{\Omega_m^0}{a^3} + \frac{\Omega_r^0}{a^4} + \Omega_\Lambda^0\right), \tag{22}$$

with the usual definitions of the dimensionless densities $\Omega_m^0, \Omega_\Lambda^0, \Omega_r^0, \Omega_k^0$. Here the index 0 refers to the present values, and we have set $a_0 = 1$. Solving the resulting equality order by order in the cosmological time, we obtain values for the dimensionless densities $\Omega_m^0, \Omega_\Lambda^0$ and the couplings $\beta, \gamma$, in terms of the purely kinematic observables $H_0, q_0, j_0, s_0$, the geometric observable $\Omega_k^0$, and[2] $\Omega_r^0$.

---

[2] Even if $\Omega_r^0$ does not have a geometric origin, we included it as an input parameter to avoid going beyond

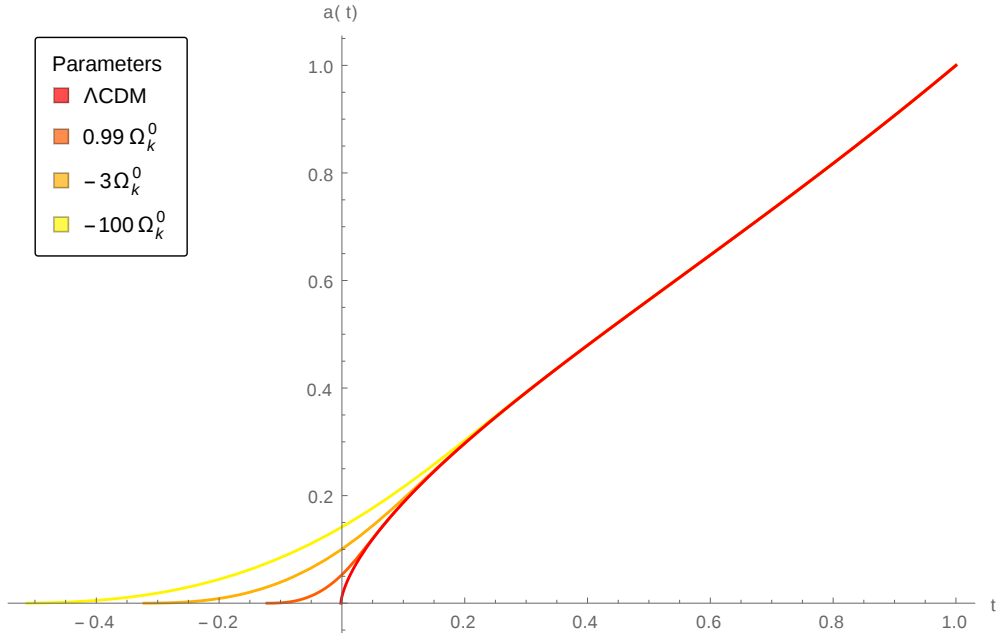

FIG. 1: The scale factor as a function of time for the values of the kinematic parameters consistent with current observations $H_0 = 0.955\ \mathcal{U}^{-1}, q_0 = -0.537, j_0 = 1, s_0 = -0.390$, for different choices of the curvature component within the observational bounds $\Omega_k^0 = 0 \pm 5 \times 10^{-3}$. Here $\mathcal{U}$ stands for the age of the Universe in the $\Lambda$CDM model. The depicted inflationary histories have a number of $e$-folds $\simeq 70$ since Planck era, that decreases as we move $\Omega_k^0$ closer to the $\Lambda$CDM value.

The resulting cosmological histories for values of $H_0, q_0, j_0, s_0$, consistent with current observations and for different choices of the curvature component $\Omega_k^0$, are shown in Figure 1. Varying $\Omega_k^0$ within the present observational bounds, we obtain cosmologies with a inflationary era followed by a radiation/matter dominated epoch and late time acceleration. Defining the beginning of the classical inflationary evolution as the time when radiation density matches Planck's density, $\Omega_k^0$ can be chosen so that the resulting inflationary era has a number of $e$-folds $\gtrsim 60$. Regarding late time acceleration, it is important to stress that it is driven by the dressed cosmological constant $\Lambda$, solution of $P[\Lambda/3] = \Lambda_0/3$. It is interesting to notice that $P'[\Lambda/3] = 1/(4\kappa_{eff})$, as defined in (14), implying that the acceleration of such de Sitter phase is well defined as long as the effective Newton's constant remains finite.

---

the fourth order in the expansion of the scale factor. This is of course not necessary in principle, including a fifth order in the expansion (21), $\Omega_r^0$ would be obtained as an output.

## 2. Cosmological inhomogeneities

In order to study cosmological inhomogeneities, we limit ourselves to the cubic case setting $\gamma = 0$ in what follows. We rewrite the flat FLRW background in conformal time $\tau$, and perturb it away from homogeneity with a scalar fluctuation, as

$$ds^2 = a^2(\tau)\left(-(1+2\Phi)d\tau^2 + (1-2\Psi)d\vec{x}^2\right) \qquad (23)$$

Defining $\mathcal{H} = a'/a$ with a prime ( )$'$ to represent the conformal time derivative $\partial_\tau($ ), the non-vanishing components of the linearized gravitational tensor read

$$E^\tau{}_\tau = \frac{2}{a^2}\left(3\mathcal{H}^2\Phi + 3\mathcal{H}\Psi' - \nabla^2\Psi\right) + \frac{48\beta\kappa^2}{a^6}\left(\left(\mathcal{H}' - \mathcal{H}^2\right)\nabla^4\Psi + 2\mathcal{H}^4\left(3\mathcal{H}^2\Phi + 3\mathcal{H}\Psi' - \nabla^2\Psi\right)\right) \qquad (24)$$

$$E^\tau{}_i = -\frac{2}{a^2}\left(\mathcal{H}\partial_i\Phi + \partial_i\Psi'\right) + \frac{48\beta\kappa^2}{a^6}\left(\left(\mathcal{H}'' - 3(\mathcal{H}' - \mathcal{H}^2)\mathcal{H}\right)\partial_i\nabla^2\Psi - 2\mathcal{H}^4\left(\mathcal{H}\partial_i\Phi + \partial_i\Psi'\right)\right.$$
$$\left. + \left(\mathcal{H}' - \mathcal{H}^2\right)\left(\mathcal{H}\partial_i\nabla^2\Phi + \partial_i\nabla^2\Psi'\right)\right) \qquad (25)$$

while for $i \neq j$ we have

$$E^i{}_j\big|_{i\neq j} = \frac{1}{a^2}\partial_i\partial^j\left(\Psi - \Phi\right) + \frac{24\kappa^2\beta}{a^6}\left(6\left(2\mathcal{H}\mathcal{H}' - \mathcal{H}''\right)\partial_i\partial^j\Psi'\right.$$
$$+ (\mathcal{H}' - \mathcal{H}^2)\left(2\partial_i\partial^j\nabla^2\Psi + \partial_i\partial^j\nabla^2\Phi - 3\partial_i\partial^j\Psi'' + 3\mathcal{H}\left(3\partial_i\partial^j\Psi' - \partial_i\partial^j\Phi'\right)\right)$$
$$\left. - \left(37\mathcal{H}^2\mathcal{H}' + 3\mathcal{H}^4 - 3\mathcal{H}\mathcal{H}'' + 6\mathcal{H}'''\right)\partial_i\partial^j\Psi + \left(15\mathcal{H}^2\mathcal{H}' - 8\mathcal{H}^4 - 3\mathcal{H}\mathcal{H}'' - 3\mathcal{H}'^2\right)\partial_i\partial^j\Phi\right). \qquad (26)$$

Finally, the diagonal equations along the Cartesian coordinates $yy$ and $zz$ can be obtained by respectively replacing $x \to y$ and $x \to z$ in the following $xx$ component

$$E^x{}_x = \frac{1}{a^2}\left(\left(\nabla^2 - \partial_x^2\right)\left(\Phi - \Psi\right) + 2\Psi'' + 4\left(\mathcal{H}' - \mathcal{H}^2\right)\Phi + \mathcal{H}\left(4\Psi' + 2\Phi'\right)\right)$$
$$+ \frac{12\kappa^2\beta}{a^6}\left[8\mathcal{H}^5\Phi' - 4\mathcal{H}^2\nabla^2\Psi'' - \mathcal{H}^3\Psi'\left(19\mathcal{H}^2 - 29\mathcal{H}'\right) - 8\mathcal{H}^4\left(3\left(\mathcal{H}^2 - 2\mathcal{H}'\right)\Phi - \Psi''\right)\right.$$
$$+ 2\left(\nabla^2\Psi' - 3\partial_x^2\Psi'\right)\left(3\mathcal{H}^3 - 7\mathcal{H}\mathcal{H}' + 2\mathcal{H}''\right) - 2\partial_x^2\Psi\left(19\mathcal{H}^2\mathcal{H}' - 15\mathcal{H}\mathcal{H}'' - 9\mathcal{H}'^2 + 3\mathcal{H}'''\right)$$
$$+ 2\left(\mathcal{H}^2 - \mathcal{H}'\right)\left(\nabla^4\Phi + 2\nabla^4\Psi - \mathcal{H}\nabla^2\Phi' + \nabla^2\Phi'' - \partial_x^2\nabla^2\Phi - 3\partial_x^2\nabla^2\Psi + 3\mathcal{H}\partial_x^2\Phi' + 3\partial_x^2\Psi''\right)$$
$$+ 2\nabla^2\Psi\left(4\mathcal{H}^4 + \mathcal{H}^2\mathcal{H}' - 5\mathcal{H}\mathcal{H}'' - 3\mathcal{H}'^2 + \mathcal{H}'''\right) + 2\nabla^2\Phi\left(4\mathcal{H}^4 - 5\mathcal{H}^2\mathcal{H}' + \mathcal{H}\mathcal{H}'' + \mathcal{H}'^2\right)$$
$$\left. - 2\partial_x^2\Phi\left(8\mathcal{H}^4 - 15\mathcal{H}^2\mathcal{H}' + 3\mathcal{H}\mathcal{H}'' + 3\mathcal{H}'^2\right)\right] \qquad (27)$$

Remarkably, since the components of the gravitational tensor are second order in time, the same is true for the equations of motion for the perturbations. Moreover, this property dissapears if any of the relative coefficients in the cubic action is moved away from the four-dimensional Einsteinian point. They are however, higher order in spatial derivatives.

## IV. CONCLUSIONS

In this note, we proved that the five-dimensional cubic quasitopological gravity of [20], when written in four dimensions, matches the four-dimensional cubic Einsteinian gravitational theory proposed in [17]. Furthermore, we constructed the quartic generalization of such model, as well as an example of a quintic contribution with the same properties.

As in the cubic case, the quatic theory theory gives rise to second order equations of motion in a FLRW ansatz, with purely geometric inflationary solution, and a third order equation with a trivial first integral in a static spherically symmetric anzatz. For kinematic parameters and spatial curvature within the observational ranges, the number of $e$-folds since Planck era is $\gtrsim 60$. Remarkably, such properties show up in the same region of parameters in which small black holes are thermodynamically stable, thus providing a good candidate for the dark matter component of the Universe. We also wrote the equations of motion for the scalar fluctuations on the cubic case, showing the remarkable property that they are second order in time.

It is beyond of the scope of this paper to find the strongest phenomenological constraints on the parameters of the model. Nevertheless, we think that cosmological models based on this kind of well behaved higher curvature theories are a promising phenomenological ground, and deserve more research.

## V. AKNOWLEDGEMENTS

The authors want to thank José Edelstein for enlightening comments, in particular regarding the sign of the cubic coupling giving rise to healthy cosmologies as oppsite to that allowing the existence of black holes. This work was partially suported by CONICET (Argentina) grant number PIP-2017-1109 and CONICYT (Chile) PAI80160018, Stays in Chile of International Excellence Researchers (N.G.), by CONICYT (Chile) grant numbers 1181047, 1150907 (J.O.) and 11170274 (A.C.).

## Appendix A: Quartic construction

As stated in section III, we constrain the quartic combination in order to fulfil

- Absense of ghost around the maximally symmetric background.

- Second order equations of motion on a homogeneous and isotropic cosmology.

- Third order equations of motion with a trivial first integral on a static, spherically symmetric ansatz compatible with $-g_{tt}g_{rr} = 1$.

These restrictions lead to the following constraints

$$c_{21} = \frac{49c_1}{30} - \frac{2c_{10}}{3} + 3c_{11} - \frac{2c_{13}}{3} - \frac{5c_{14}}{6} - \frac{c_{15}}{3} - c_{16} - 2c_{18} + \frac{5c_2}{3} - \frac{c_{22}}{2} + \frac{2c_3}{3} + \frac{18c_4}{5}$$
$$+ \frac{36c_5}{5} + \frac{58c_6}{15} + \frac{104c_7}{5} + \frac{5c_8}{6}$$

$$c_{24} = \frac{73c_1}{30} - \frac{c_{10}}{3} + 4c_{11} - c_{13} - \frac{c_{14}}{6} - \frac{2c_{15}}{3} - \frac{4c_{16}}{3} - 2c_18 + 3c_2 - \frac{7c_{22}}{6} + c_3 + \frac{58c_4}{15}$$
$$+ \frac{116c_5}{15} + \frac{86c_6}{15} + \frac{424c_7}{15} + \frac{7c_8}{6} - \frac{2c_9}{3},$$

$$c_{12} = -\frac{19c_1}{60} - \frac{c_{11}}{2} - \frac{c_2}{2} - \frac{c_3}{12} - \frac{4c_4}{5} - \frac{8c_5}{5} - \frac{14c_6}{15} - \frac{56c_7}{15} - \frac{c_8}{8} - \frac{c_9}{4},$$

$$c_{17} = -\frac{23c_1}{30} - \frac{11c_{10}}{12} - \frac{c_{13}}{6} - \frac{c_{14}}{3} - \frac{c_{15}}{12} - \frac{c_{16}}{4} - \frac{4c_2}{3} - \frac{c_3}{12} - \frac{11c_4}{5} - \frac{22c_5}{5} - \frac{41c_6}{15}$$
$$- \frac{28c_7}{5} - \frac{c_8}{24} - c_9,$$

$$c_{19} = \frac{11c_1}{30} + \frac{c_{10}}{6} + \frac{3c_{11}}{8} - \frac{c_{13}}{48} + \frac{c_{14}}{12} - \frac{c_{15}}{24} - \frac{c_18}{4} + \frac{7c_2}{12} + \frac{c_3}{12} + \frac{9c_4}{10} + \frac{9c_5}{5} + \frac{17c_6}{15}$$
$$+ \frac{16c_7}{5} + \frac{5c_8}{48} + \frac{c_9}{4},$$

$$c_{20} = \frac{3c_1}{5} - c_{13} + c_{14} - 2c_{15} + 2c_2 + c_{22} - \frac{4c_4}{5} - \frac{8c_5}{5} + \frac{12c_6}{5} + \frac{48c_7}{5} - 2c_9,$$

$$c_{23} = \frac{17c_1}{15} + \frac{4c_{10}}{3} + \frac{c_{13}}{12} + \frac{c_{14}}{6} + \frac{c_{15}}{6} + \frac{5c_2}{3} - \frac{c_{22}}{2} + \frac{c_3}{6} + \frac{18c_4}{5} + \frac{36c_5}{5} + \frac{58c_6}{15}$$
$$+ \frac{24c_7}{5} + \frac{c_8}{12} + \frac{3c_9}{2},$$

$$c_{25} = -\frac{97c_1}{30} - \frac{c_{10}}{12} - \frac{9c_{11}}{2} + \frac{25c_{13}}{24} + \frac{c_{14}}{12} + \frac{5c_{15}}{6} + \frac{5c_{16}}{4} + 2c_18 - \frac{25c_2}{6} + \frac{3c_{22}}{4} - \frac{7c_3}{6}$$
$$- \frac{29c_4}{5} - \frac{58c_5}{5} - \frac{124c_6}{15} - \frac{172c_7}{5} - \frac{4c_8}{3} + \frac{c_9}{4},$$

$$c_{26} = \frac{9c_1}{20} + \frac{5c_{11}}{8} - \frac{7c_{13}}{48} - \frac{c_{15}}{8} - \frac{c_{16}}{6} - \frac{c_18}{4} + \frac{7c_2}{12} - \frac{c_{22}}{12} + \frac{c_3}{6} + \frac{11c_4}{15} + \frac{22c_5}{15} + \frac{17c_6}{15}$$
$$+ \frac{73c_7}{15} + \frac{3c_8}{16} - \frac{c_9}{12}$$

**Appendix B: Quintic construction**

In order to construct a quintic four-dimensional Einsteinian theory with the desired properties, we write a generic combination quintic invariants as

$$d_1 R \left(R_{ab}R^{ab}\right)^2 + d_2 RR_b{}^a R_c{}^b R_d{}^c R_a{}^d + d_3 RR_b{}^a R_d{}^b R_e{}^c R_{ac}{}^{de} + d_4 R_c{}^a R_d{}^b R_{ef}{}^{cd} R_{ab}{}^{ef}$$

$$+ d_5 RR_c{}^a R_e{}^b R_{af}{}^{cd} R_{bd}{}^{ef} + d_6 RR_c^a R_e^b RR_{bf}{}^{cd} R_{ad}{}^{ef} + d_7 RR_b^a R_{ad}{}^{bc} R_{fg}{}^{de} R_{ce}{}^{fg}$$

$$+ d_8 RR_b^a R_{de}{}^{bc} R_{fg}{}^{de} R_{ac}{}^{fg} + d_9 RR_b^a R_{df}{}^{bc} R_{ag}{}^{de} R_{ce}{}^{fg} + d_{10} RR_{cd}{}^{ab} R_{ae}{}^{cd} R_{gh}{}^{ef} R_{bf}{}^{gh}$$

$$+ d_{11} RR_{cd}{}^{ab} R_{ef}{}^{cd} R_{gh}{}^{ef} R_{ab}{}^{gh} + d_{12} RR_{ce}{}^{ab} R_{ag}{}^{cd} R_{bh}{}^{ef} R_{df}{}^{gh} + d_{13} RR_{ce}{}^{ab} R_{ag}{}^{cd} R_{dh}{}^{ef} R_{bf}{}^{gh}$$

$$+ d_{14} R_b^a R_a^b R_d^c R_e^d R_c^e + d_{15} R_b^a R_a^b R_e^c R_f^d R_{cd}{}^{ef} + d_{16} R_b^a R_e^b R_d^c R_f^d R_{ac}{}^{ef} + d_{17} R_b^a R_a^b R_d^c R_{fg}{}^{de} R_{ce}{}^{fg}$$

$$+ d_{18} R_b^a R_d^b R_f^c R_{ag}{}^{de} R_{ce}{}^{fg} + d_{19} R_b^a R_a^b R_{ef}{}^{cd} R_{gh}{}^{ef} R_{cd}{}^{gh} + d_{20} R_b^a R_a^b R_{eg}{}^{cd} R_{ch}{}^{ef} R_{df}{}^{gh}$$

$$+ d_{21} R_b^a R_c^b R_{ae}{}^{cd} R_{gh}{}^{ef} R_{df}{}^{gh} + d_{22} R_c^a R_g^b R_{ae}{}^{cd} R_{dh}{}^{ef} R_{bf}{}^{gh} + d_{23} R_b^a R_{ad}{}^{bc} R_{cf}{}^{de} R_{hi}{}^{fg} R_{eg}{}^{hi}$$

$$+ d_{24} R_b^a R_{ad}{}^{bc} R_{fg}{}^{de} R_{hi}{}^{fg} R_{ce}{}^{hi} + d_{25} R_{cd}{}^{ab} R_{ae}{}^{cd} R_{bg}{}^{ef} R_{ij}{}^{gh} R_{fh}{}^{ij}$$

$$+ d_{26} R_{ce}{}^{ab} R_{fg}{}^{cd} R_{hi}{}^{ef} R_{aj}{}^{gh} R_{bd}{}^{ij}.$$

On the resulting dynamics, we impose the following constraints

- When considered as a five dimensional theory, it must satisfy the quasitopological properties (we impose this condition in order to reduce the number of invariants that may appear).

  - Being ghost-free around the maximally symmetric background.
  - Providing first order equations of motion in a spherically symmetric ansatz, *i.e.* having a Birkhoff theorem.

- When considered as a four dimensional theory, it must satisfy the Einsteinian properties

  - Being ghost-free around the maximally symmetric background.
  - Leading to second order equations of motion on a FLRW ansatz.
  - Leading to third order equations of motion with a trivial first integral on a static, spherically symmetric ansatz, fulfilling $-g_{tt}g_{rr} = 1$.

With these conditions, we obtain the following restrictions on the $d_i$'s

$$d_1 = \frac{81847}{115648}\xi_5, \quad d_2 = -\frac{69959}{173472}\xi_5, \quad d_3 = -\frac{58427}{10842}\xi_5, \quad d_4 = \frac{38813}{21684}\xi_5,$$

$$d_5 = -\frac{535655}{86736}\xi_5, \quad d_6 = \frac{214553}{28912}\xi_5, \quad d_7 = -\frac{167065}{86736}\xi_5, \quad d_8 = \frac{69455}{14456}\xi_5,$$

$$d_9 = \frac{23135}{834}\xi_5, \quad d_{10} = \frac{198411}{115648}\xi_5, \quad d_{11} = \frac{60085}{57824}\xi_5, \quad d_{12} = -\frac{15267}{28912}\xi_5,$$

$$d_{13} = -\frac{277981}{28912}\xi_5, \quad d_{14} = -\frac{27623}{43368}\xi_5, \quad d_{15} = \frac{190787}{43368}\xi_5, \quad d_{16} = -\frac{102481}{21684}\xi_5,$$

$$d_{17} = -\frac{12868}{5421}\xi_5, \quad d_{18} = -\frac{814}{1807}\xi_5, \quad d_{19} = -\frac{977503}{173472}\xi_5, \quad d_{20} = -\frac{1794913}{86736}\xi_5,$$

$$d_{21} = \frac{4555}{14456}\xi_5, \quad d_{22} = \frac{122137}{7228}\xi_5, \quad d_{23} = \frac{54215}{14456}\xi_5, \quad d_{24} = -\frac{330845}{86736}\xi_5,$$

$$d_{25} = \frac{215}{416}\xi_5, \quad d_{26} = \xi_5$$